# Anchoring-mediated stick-slip winding of cholesteric liquid crystals

Weichao Zheng[1],*

[1]*Department of Chemistry, Physical and Theoretical Chemistry Laboratory, University of Oxford, Oxford OX1 3QZ, United Kingdom*

The stick-slip phenomenon widely exists in contact mechanics, from the macroscale to the nanoscale. During cholesteric-nematic unwinding by external fields, there is controversy regarding the role of planar surface anchoring, which may induce discontinuous stick-slip behaviors despite the well-known continuous transitions observed in past experiments. Here, we observe three regimes, namely constrained, stick-slip, and sliding-slip, under mechanical winding with different anchoring conditions, and measure the responded forces by the Surface Force Balance. These behaviors result from a balance of cholesteric elastic torque and surface torque, reminiscent of the slip morphology on frictional substrates [T. G. Sano *et al*., Phys. Rev. Lett. **118**, 178001 (2017)], and provide evidence of dynamics in static rotational friction.

In broad soft matter areas, including turbulence [1], micro/nanofluidics [2], and yield stress materials [3], boundary conditions are important for material properties and performance. Similarly, in liquid crystals, surface anchoring also plays a crucial role in the order parameter, the temperature of the nematic-isotropic phase transition [4], and the response of molecules to external fields [5], especially in confined geometries such as liquid crystal displays. With strong anchoring, there exists a critical threshold voltage that orients the nematic molecules, called the Fréedericksz transition [6], below which molecules are still. However, there is a debate about whether planar anchoring affects the cholesteric-nematic unwinding transition by external fields. Decades ago, it was predicted [5,7] and proven [8-11] that magnetic or electric fields can continuously unwind cholesterics to nematics, but the situation with different planar anchoring conditions was not explicitly addressed by experiments. Some studies [7,12,13] suggested that the continuous cholesteric-nematic transition is only applicable for bulk samples in which the surface anchoring is negligible. By varying the anchoring strength in confinement, rich behaviors, such as stick-slip or step-wise transitions, were predicted to happen under external stimuli [7,13-27], such as temperature, light, stress, magnetic and electric fields. Particularly, a recent study [28] reported that if the easy axis on one surface rotates, chiral nematics may show three regimes, including free twist, stick-slip, and constrained winding, as a balance of twist elastic torque and surface torque [20,28-30]. Although some evidence of discontinuous transitions has been presented [12,31-37], different mechanisms were still discussed, probably due to the experimental precision and the complexity of surface anchoring, and none of the models could be directly applied to explain the observations in this work.

Here, we use the Surface Force Balance (SFB) to simultaneously measure the optical and mechanical responses of cholesterics along the helical axis under various boundary conditions. Desiccated cholesterics were confined between two freshly-cleaved muscovite mica surfaces that were glued onto crossed cylinders. In the beginning, a strong planar anchoring was obtained, but anchoring strength decayed over time mainly due to the adsorption of water from the ambient environment [38,39]. Therefore, three different regimes were observed during experiments, resulting from the decayed frictional surface torque. Furthermore, the hysteresis of twist transitions was observed during the retraction and approach of surfaces in all three regimes.

*Three regimes*.--Cholesteric layers, with a layer thickness of half-pitch $p = 122$ nm, were compressed in the SFB [40] [Fig. 1(a)] with a cylinder radius of $R$. With time evolution, three regimes of the measured force profiles were observed [Fig. 1(b)]. In the first regime, the force generated by the constrained cholesterics initially started from zero and increased with increasing strain to 65%, peaking at 14 mN/m before the surface jumped into contact position, and all the cholesterics were squeezed out together. In the second regime, stick-slip jumps of the surface occurred after the force accumulated to 1.5 mN/m with about 30% strain, and finally, the surfaces jumped to contact. The number of jumps corresponded to five integral layers and a non-integral layer since the easy axes on mica surfaces were not parallel [40]. Sometimes, multiple-layered jumping events were observed in this regime [Fig. 1(b) and Fig. S1]. In the last regime, the surface jumped periodically with a wavelength equal to the half-pitch without a large deformation of cholesterics, and the last few layers were difficult to squeeze out, resulting in large forces. It is worth noting that there was a constant background force of around 1 mN/m in this regime.

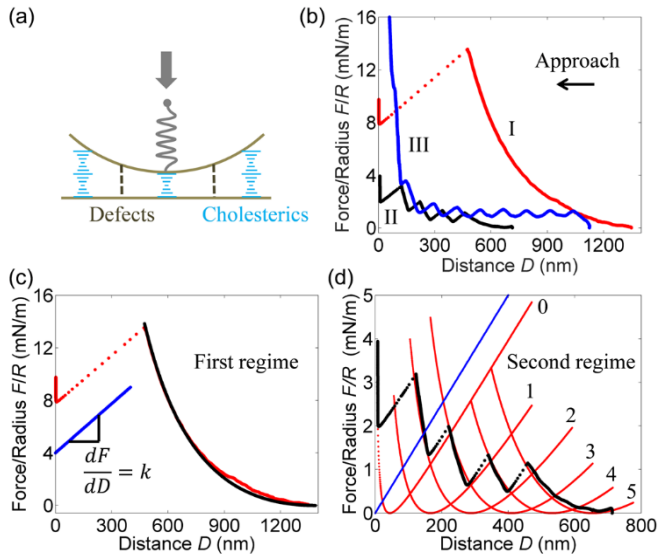

FIG. 1. Forces measured in the Surface Force Balance (SFB). (a) Schematic diagram of cholesterics confined in the crossed cylinders with radius $R$. (b) Force profiles of three regimes, i.e., I Constrained (red), II Stick-slip (black), and III Sliding-slip (blue), during the approach of the surface as the anchoring strength decreases. (c) Force profile (red) in the first regime fitted by elastic forces (black) calculated by Equation A4 with $K_{22} = 3.8$ pN. (d) Force profile (black) in the second regime fitted by elastic forces (red) calculated by Equation A4 with various integral layers (numbers) and $K_{22} = 6$ pN. The slope of the blue line in (c) and (d) is the spring constant of the cantilever spring that connects to the surface $dF/dD = k = 125$ N/m.

Fig. 1(c) shows that the match between the experiment and theory (Equation A4) is good, indicating that the anchoring strength in the first regime is strong. The slope of the jump-in is comparable to the spring constant, which manifests that the spring instability dominates the jumping process [40]. The elastic deformation almost without dissipation [41] works like an ideal spring, neglecting the effect of gravity at the micro/nanoscale. This deformation, which can last for more than one hour without dissipation if the surface stops moving [Fig. S2], is truly elastic rather than viscous.

In the second regime, the force profile can also be fitted by the harmonic elastic forces calculated using Equation A4 with various layers of cholesterics [Fig. 1(d)]. The crossing points of the harmonic forces fall on distances equal to the integral number of quarter-pitch [Fig. S3]. Notably, the slope of forces during the jump is a little larger than the theoretical one but comparable to the spring constant, indicating that the jumping process is a balance of both viscoelastic forces and spring force, which is shown in Fig. 1(d), where jumping distances are smaller than theoretical values. This deviation of jumping distances could be due to the expansion of dislocation defects that store elastic energy. The effect of defects is discussed in the Supplemental Material.

*Discussion*.--The compression ratio that cholesterics can sustain with time decreases from the first to the third regime, indicating a decrease in anchoring strength after the adsorption of water from the ambient environment [38,39]. In the third regime, surfaces are difficult to compress to contact, which supports the assumption that surfaces are changing with time. There are several reasons why a hard wall is encountered before contact. Firstly, the adsorbed water dissolves and accumulates potassium ions from mica surfaces to the contact position, which increases the electrostatic repulsion. Secondly, liquid crystal molecules grow epitaxially with time [42]. Thirdly, contaminants from the ambient air adsorb to the surface.

These three regimes emerge with the change of anchoring strength. Considering the longer timescale, more regimes might appear. For example, if the adsorbed water changes the direction of the easy axis on the mica surface [38,39], the behaviors could be different. Finally, if the mica surfaces become totally homeotropic, the pitch axis will be parallel with the surface, causing fingerprint textures and more isotropic-like optics.

*Surface torque*.--The measured forces follow the twist elastic theory very well with manual input of the twist angle, but three different regimes varying with anchoring strength are obtained, namely constrained, stick-slip, and sliding-slip. What is the mechanism that determines the critical threshold of the jump in different regimes?

When cholesterics are confined between two plates, the elastic torque is balanced by the surface torque, which includes the surface anchoring and surface viscosity [28-30]. For strong anchoring, molecules deviate very slowly from the easy axis, thus the torque from surface viscosity is negligible. While at mediate anchoring, molecules slide to a deviated angle with a larger speed, therefore, both surface anchoring and surface viscosity balance with elastic torque.

Fig. 2 (a and b) shows that there exists a threshold constant of the compression ratio, about 35% and 75% in the first and second regimes respectively, for the surface to sustain the elastic stress at a certain anchoring condition, no matter how many layers are compressed. This constant ratio manifests that there is a threshold anchoring torque $\Gamma_c$ that is analogous to the breakaway friction torque [43], a concept from rotational friction. When the anchoring is strong, the frictional torque can sustain a large elastic torque, such that cholesteric layers won't jump until the threshold is reached.

The anchoring torque in the first and second regimes is plotted in Fig. 2(c), where the first regime and second regime fall on two slopes of calculation based on Equation B3 with twist elastic constant $K_{22} = 3.8$ and 6 pN, respectively. If the force profile in Fig. 1(c) is carefully examined, one can see that the slope at small compression is actually higher than the calculation with $K_{22} = 3.8$ pN. This may be because mica surfaces on cylindrical lenses are not large enough, such that at small compression, the forces mainly generated near contact position are free from the effect of mica areas, but at the large compression, mica areas start to limit the force responses, producing smaller forces. From the fitted elastic constant, we can estimate the effective coverage of mica on lenses is 2/3.

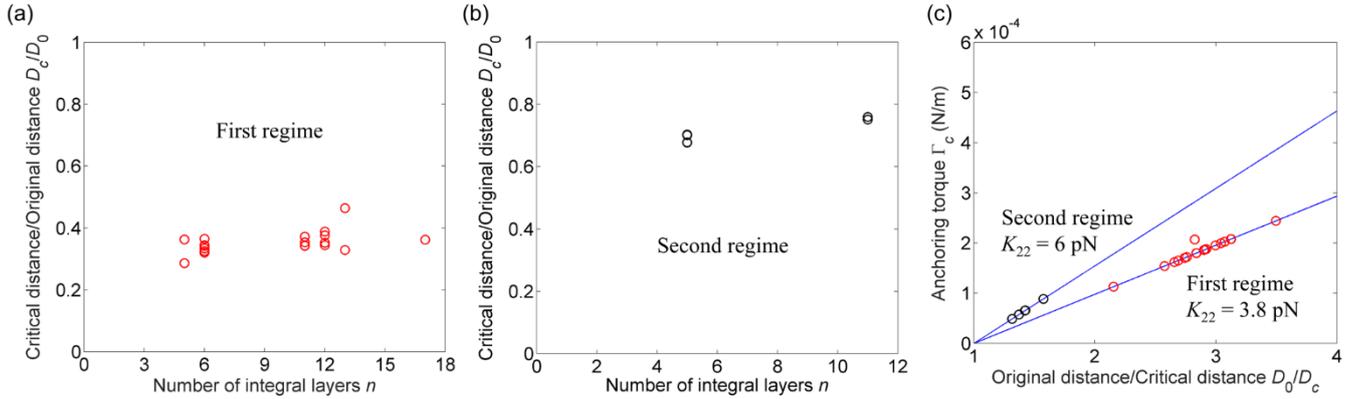

FIG. 2. Compression ratio of cholesteric layers at the critical jumping distance. (a) First regime. (b) Second regime (premier jumps). (c) The data in the first and second regimes fall on two blue lines that are theoretically calculated by Equation B3 with $K_{22} = 3.8$ and 6 pN, respectively.

From Fig. 3(a), the slope and intersection obtained from the trend line are used to calculate the anchoring strength by Equation B6. Therefore, critical anchoring torque $\Gamma_c \approx 0.23$ mN/m and anchoring strength $W \approx 0.15$ mN/m with $K_{22} = 6$ pN and half-pitch $p = 122$ nm. The deviated angle on one surface can be calculated $\frac{\Phi_0 - \Phi}{2} \approx 0.49\pi$, where $\Phi_0$ is the original twist angle and $\Phi$ is the instant twist angle, which means the molecules on each surface deviate around 90° from the easy axis at the jump threshold. Notably, no mathematical solution was found with the Rapini-Papoula potential [5], but the anchoring potential with other forms may also be feasible. For example, if the anchoring torque is $\frac{1}{2}W\frac{\Phi_0 - \Phi}{2}$ differentiated from the anchoring energy $\frac{1}{2}W(\frac{\Phi_0-\Phi}{2})^2$, the anchoring strength will be 0.3 mN/m, but the critical torque and deviated angle are independent of the form of the anchoring potential. The exact anchoring potential could be further confirmed by optical observation of the deviated angle [44]. With the obtained anchoring strength and deviated angle, the measured forces in the first regime [Fig. 1(c)] can be better fitted by the elastic force by taking into account the anchoring energy [Fig. S4(a)].

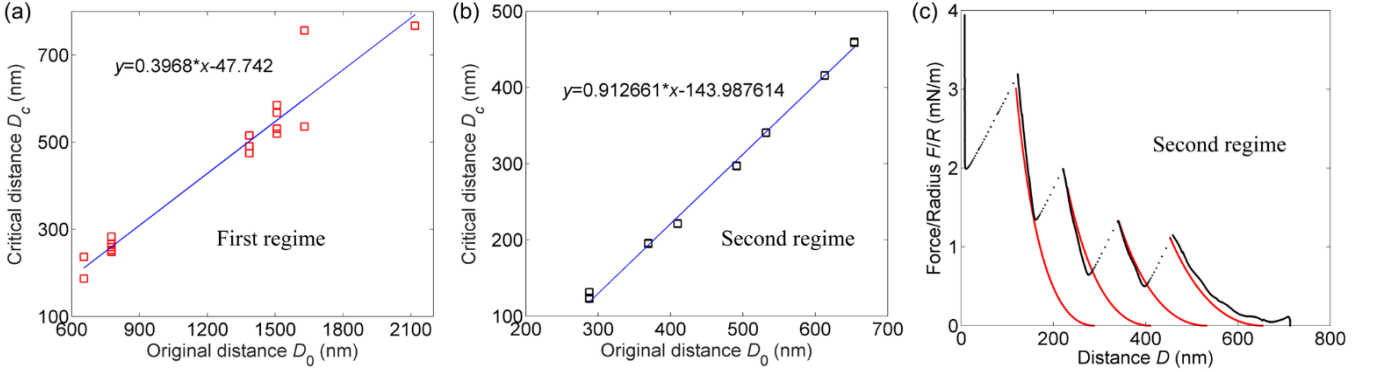

FIG. 3. Calculation of the anchoring strength. The critical jumping distance $D_c$ as a function of the original distance $D_0$ in (a) the first regime and (b) the second regime (including all the stick-slip jumps), the blue line is the linear trend line. (c) Fitting the force profile (black) in the second regime by Equations A4 and B6 with the critical surface torque and anchoring strength obtained from (b).

Similarly, the critical anchoring torque $\Gamma_c \approx 0.0148$ mN/m, the anchoring strength $W \approx 0.0073$ mN/m, and $\frac{\Phi_0-\Phi}{2} \approx 116.4°$ for the second regime are calculated from the slope and intersection in Fig. 3(b). These values are used to predict the positions where consecutive jumps occur, as shown in Fig. 3(c). The critical jumping distances fit the experimental data very well. However, the measured forces [Fig. 1(d)] are worse fitted by the elastic theory considering the anchoring energy [Fig. S4(b)]. It is as if the surface torque is correct but the composition of the torque is not a pure anchoring torque. Possibly, the surface viscosity may start to become important in this regime with medium anchoring strength. Alternatively, the 2/3 coverage of mica on the lenses may cause slip on this regime after water adsorption, since the premier critical compression ratio [Fig. 1(d)] is similar to the compression ratio where $K_{22}$ changes from a larger value to 3.8 pN in the first regime, as shown in Fig. 1(c). Last but not least, the surface torque may be balanced by the viscoelastic torque in the stick-slip regime.

In the second regime, no defects are observed stretching on the surface during either approach or retraction, indicating that the defects are in the bisector of surfaces and the polar anchoring strength [45] is larger than $2\sqrt{\frac{3}{8}K_{33}B} \approx 0.4$ mN/m, where $K_{33} = 27.5$ pN is the bend elastic constant, and $B$ is the dilation term. It seems reasonable that the azimuthal anchoring strength is one or two orders of magnitude smaller than the polar anchoring [46]. Then the polar anchoring strength in the first regime would be very large.

For weak anchoring, the anchoring torque is negligible [29]. Therefore, the elastic torque is mainly balanced by the surface viscosity. As a result, the surface viscosity can be estimated as $\gamma_s = 1.83 \times 10^{-4}$ Pa s m, and the corresponding viscous force is about 0.8 mN/s at a distance $D_0 = 1000$ nm (see Supplemental Material), which is very close to the background force in the third regime [Fig. 1(b)]. This background force may be related to the commonly observed background forces with liquid crystals in the SFA [47,48] (see Supplemental Material).

In fact, the discontinuous twist transition has been attributed to surface anchoring by most past studies [7,12-29,31-35], among which only some [20,28,29,32] used the concept of surface torque to explain the mechanism. However, many of them

[14,16-18,22-26,33,35] differentiated the anchoring energy $G$ with respect to the twist angle $\partial G/\partial \Phi$, which is actually the form of torque. The surface torque has long been adopted to describe the surface forces imposed on liquid crystals [5,49-55], but this concept does not seem to be widely used in the liquid crystal community. In a recent study [37] explaining the discontinuous transition with the energy barrier from dislocation defects, the integrating range of the equations for calculating the nucleation energy should not be the same for different layers. Therefore, the conclusions about the energy barrier were untenable.

*Hysteresis*.--Fig. 4 shows that hysteresis of the twist angle between retraction and approach exists in all three regimes and decreases with time evolution. The twist angles can be further confirmed by the 4x4 matrix simulation [40,56] [Fig. S5]. In particular, multiple-layer jumping events occur during both approach and retraction [Fig. 4 (a and c)]. Fig. 4(f) shows that the retraction profiles are the same in the first two regimes and a delayed jump resulting from the viscous stretch on the surface (see Movie S1 and Supplemental Material), is observed in the third regime. Notably, the twist angle profile during the approach in the third regime is coincident with the profile during retraction in the first two regimes, as shown in Fig. 4(d), indicating negligible anchoring torque during the approach. Most of the jumping points occur at integral quarter-pitch distances, but more uncertainties are observed at small distances.

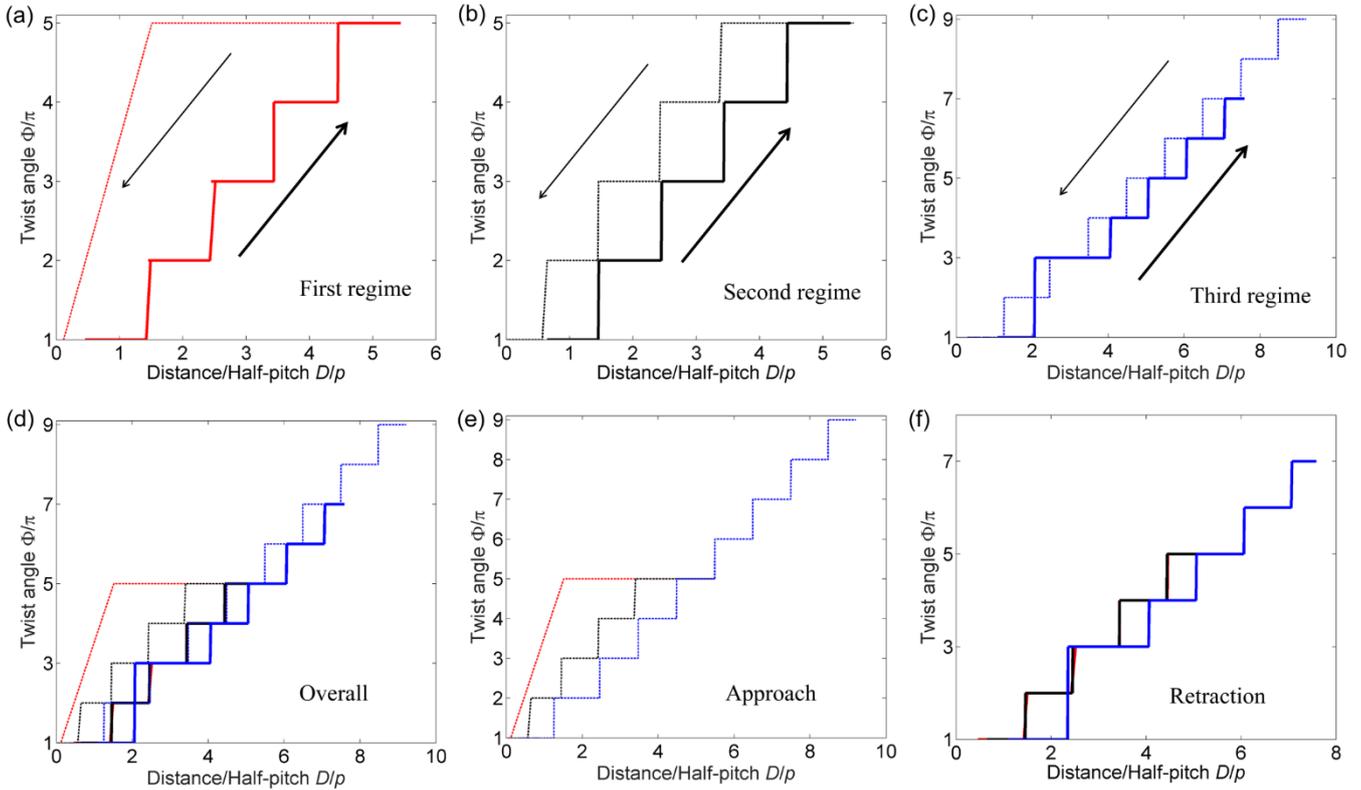

FIG. 4. Hysteresis of the twist angle in three regimes. The non-integral layer has been deducted to eliminate the difference of easy axes among different experiments. (a) First regime. The twist angle during the jump process is assumed to keep a constant compression ratio but decrease the total twist angle. (b) Second regime. (c) Third regime. The deviation of the anchoring angle is ignored in all three regimes. (d) Three regimes. (e) Approach profiles of three regimes. (f) Retraction profiles of three regimes. Thin and thick lines following the direction of the arrow are approach and retraction profiles, respectively. The red, black, and blue lines denote the three regimes, respectively.

During retraction and approach, the mechanical responses are very different, showing hysteresis, which can be understood by analogy to fracture in solid materials during tension and compression. However, given the complexity of the analogy from solids to liquid crystals, this topic will be discussed in a separate paper.

In conclusion, three regimes were observed in cholesterics during mechanical compression in the SFB. The elastic torque of cholesterics is balanced by the surface torque, which consists of anchoring torque and viscous torque. In the constrained regime with strong anchoring, the anchoring torque dominates, while the viscous torque dominates in the sliding-slip regime with weak anchoring. In the stick-slip regime, both anchoring torque and viscous torque, as well as mica coverage are possible to affect the stick-slip. This study provides a new method based on the critical surface torque to measure strong anchoring strength and deviation. The surface torque, i.e., frictional torque in rotational friction, elucidates the dynamics of static friction [57,58], as evidenced by the deviation of the anchoring angle and the hysteresis of the twist angle. This study sheds light on the understanding of boundary effects on permeative flows [59,60], friction, yield stress materials [3,61], adhesions, and biomechanics.

W.Z. is very grateful to S. Perkin for her generous help and insightful guidance on the project. S.P. suggested using the torque balance to analyze the data and the harmonic elastic potential to demonstrate the second regime. S.P. also contributed to the design of experiments and the analysis of several figures. W.Z. thanks R. Lhermerout for his derivation of equations calculating the anchoring strength, critical torque, and anchoring deviation. W.Z. is very grateful to C. S. Perez-Martinez for her assistance with some experiments. W.Z. acknowledges J. Hallett and B. Zappone for helpful discussions. Part of the work has been presented in the Ph.D. thesis titled "Optical and mechanical responses of liquid crystals under confinement (2020)". This work was supported by the European Research Council (under Grant Nos. ERC-2015-StG-676861 and 674979-NANOTRANS).

## APPENDIX A: FREE ENERGY

In cholesterics, the free energy per unit area is formed by the elastic energy and the anchoring energy from both surfaces. The anchoring potential is not a well-defined term; thus, a general parabolic form is given below,

$$G = \int_0^D \frac{1}{2} K_{22}(\phi' - q_0)^2 dD + W\left(\frac{\Phi_0 - \Phi}{2}\right)^2, \quad (A1)$$

where $D$ is the closest surface separation between two crossed cylinders, $K_{22}$ is the twist elastic constant, $\phi' = \frac{\partial \Phi}{\partial D} = \frac{\Phi}{D}$ is the molecular rotation rate at a distance $D$ with a total twist angle $\Phi$, which is constant for a uniform sample, $W$ is the anchoring strength, $\Phi_0$ is the original twist angle, and $q_0$ is the natural molecular rotation rate at relaxation. By ignoring the anchoring energy with a strong boundary, the free energy becomes,

$$G = \frac{1}{2} K_{22}\left(\frac{\Phi}{D} - q_0\right)^2 D. \quad (A2)$$

With a strong anchoring, the twist angle $\Phi \approx \Phi_0 = q_0 D_0^n$ keeps the original rotation rate at a starting distance $D_0^n$ with $n$ layers. Thus, the free energy $G^n$ and the generated force $F$ with Derjaguin approximation are written as,

$$G^n = \frac{1}{2} K_{22} q_0^2 \frac{(D_0^n - D)^2}{D}, \quad (A3)$$

$$F = 2\pi R G^n = \pi R K_{22} q_0^2 \frac{(D_0^n - D)^2}{D}. \quad (A4)$$

## APPENDIX B: SURFACE TORQUE

The torque balance can be written as,

$$K_{22}\left(\frac{\partial \Phi}{\partial D} - q_0\right) = W \frac{\Phi_0 - \Phi}{2} - \gamma_s \frac{\partial \Phi}{\partial t}, \quad (B1)$$

where $W$ is the anchoring strength, $\gamma_s$ is the surface viscosity, and $t$ is the time.

With strong anchoring, the surface viscosity and anchoring deviation are neglected here. The elastic torque $\Gamma_e$ is mainly balanced by the anchoring torque $\Gamma_a$,

$$\Gamma_e = K_{22}\left(\frac{\Phi}{D} - q_0\right) = \Gamma_a = W \frac{\Phi_0 - \Phi}{2}, \quad (B2)$$

$$K_{22} q_0 \left(\frac{D_0}{D} - 1\right) = W \frac{\Phi_0 - \Phi}{2}. \quad (B3)$$

For a more rigorous calculation that considers anchoring deviations, the surface distance $D$ in Equation B2 is calculated as,

$$D = \frac{K_{22} \Phi}{K_{22} q_0 + W \frac{\Phi_0 - \Phi}{2}}. \quad (B4)$$

At the critical torque threshold $\Gamma_c$, the critical twist angle $\Phi_c$ and critical surface distance $D_c$ are calculated below,

$$\Phi_c = \Phi_0 - \frac{2\Gamma_c}{W}, \tag{B5}$$

$$D_c = \frac{K_{22}(\Phi_0 - \frac{2\Gamma_c}{W})}{K_{22}q_0 + \Gamma_c} = \frac{D_0 - \frac{2\Gamma_c}{q_0 W}}{1 + \frac{\Gamma_c}{K_{22}q_0}}. \tag{B6}$$

*zwhich@outlook.com

# Supplemental Material

## I. FORCE MEASUREMENTS

The speed of the motor in the SFB is usually calibrated with a baseline at large distances without encountering any forces, although there is a control signal roughly showing the speed. However, with cholesterics inside the SFB, there is always an elastic background that disturbs the precise calibration of the speed. Therefore, the speed of the motor was estimated within reasonable ranges in the first regime (mostly 1.6-1.8 nm/s for the motor signal 0.02 VE) and the second regime (1.3-1.8 nm/s for the motor signal 0.02 VE) by differentiating the distance profile. Then the obtained force profiles were compared with the theoretical forces calculated using Equation A4 with the intrinsic twist elastic constant of cholesterics, $K_{22} = 6.18$ pN, reported elsewhere [1]. It is worth noting that the main argument in this work, i.e., surface torque, is independent of the force calibration.

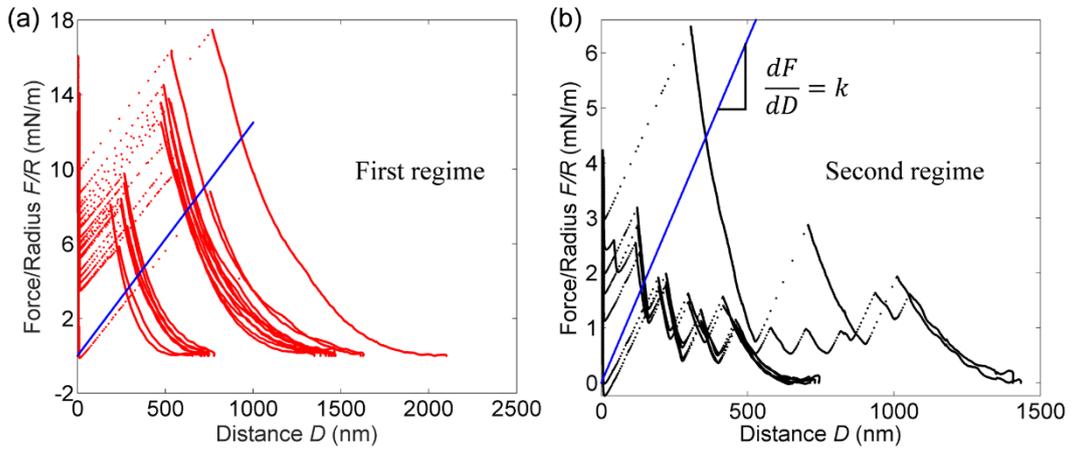

FIG. S1 Force measurements in the SFB. (a) First regime. (b) Second regime. The slope of the blue line in (a) and (b) is the spring constant of the cantilever spring that connects to the surface, $k = 125$ N/m.

Fig. S2(a) shows force responses as a function of time in the first regime with 50% strain but before reaching the critical jumping distance. At $t \approx 20$ mins, the motor stopped and re-approached at $t \approx 90$ mins, between which the surface distance was sustained by the cholesterics with a deviation of less than 40 nm mainly due to mechanical and thermal drifts, as shown in Fig. S2(b). After re-approach, cholesterics were further compressed before all layers were squeezed out, and the surfaces jumped into contact. The long-time study demonstrates that the surfaces experienced elastic forces rather than dissipative viscous forces.

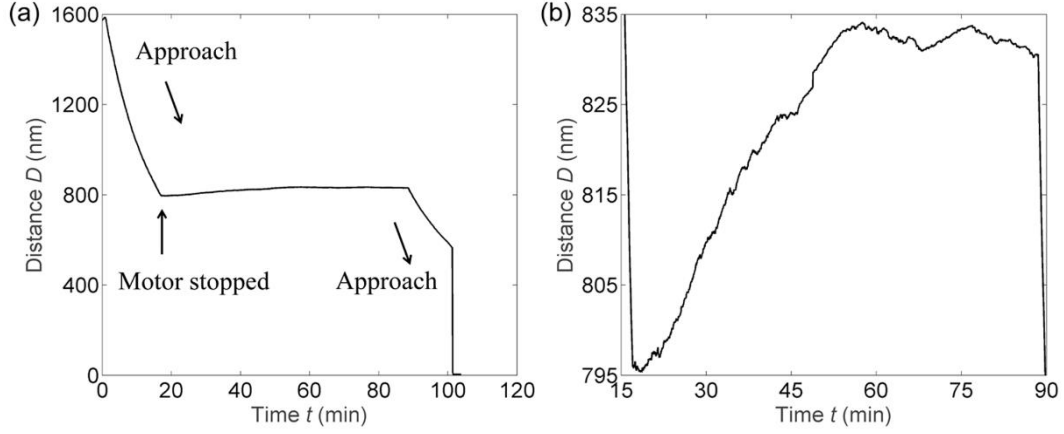

FIG. S2 Elastic response of cholesterics. (a) The approach and stop of the surface. The motor stopped at $t \approx 20$ mins and restarted at $t \approx 90$ mins during the approach. (b) Zoom-in view of the distance profile.

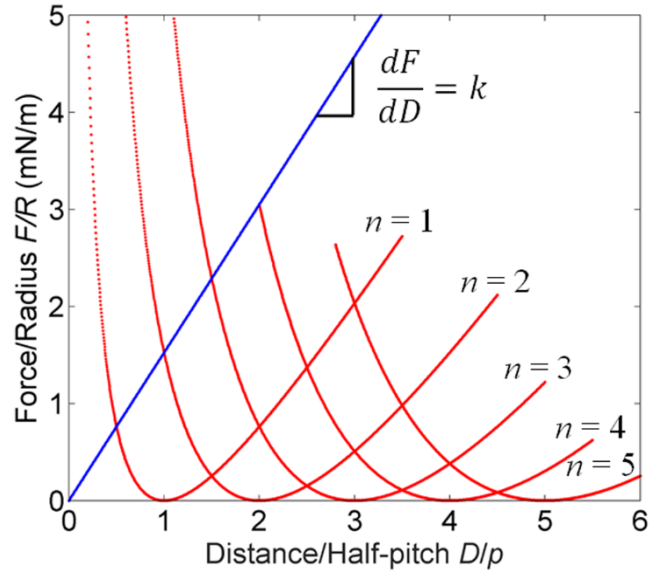

FIG. S3 Harmonic elastic forces calculated by Equation A4 with various integral layers $n$ and $K_{22} = 6$ pN. The slope of the blue line is the spring constant of the cantilever spring $k = 125$ N/m.

## II. EFFECT OF DEFECT ENERGY

When the crossed cylinders are close enough, the geometry is similar to a sphere with a radius $R$ approaching a flat surface. Therefore, the height $h$ at a radius $r$ from the contact point is calculated by,

$$h - D = \frac{r^2}{2R}, \tag{S1}$$

where $D$ is the closest surface separation. All circular defects confined in the SFB stay between integral cholesteric layers. If, at the contact position (with 0 layers confined), the radius of the innermost dislocation defect with a height $p/2$ ($p$ is the half-pitch) is defined as $r_{0,1}$, the defect energy is the sum of all defects $2\pi\mu \sum_{i=1}^{n} r_{0,i}$, with the line tension $\mu \approx K_{22}$. Similarly, with $m$ relaxed layers confined in the SFB, the defect energy can be calculated as $2\pi\mu \sum_{i=m+1}^{n} r_{m,i}$. Using Equation S1, the relationship $r_{m,m+1} = r_{0,1} = \sqrt{pR}$ is found. As a result, the difference of the defect energy between $m$ and 0 layers at the relaxed state is $2\pi\mu \sum_{i=n-m+1}^{n} r_{0,i}$.

In the first regime, if 10 layers are compressed from the distance $D = 10p$ to a critical jumping distance $D = 4p$, the twist elastic energy can be calculated by integrating Equation A4, $E = 2.54 \times 10^{-11}$ N·m. At distance $D = 4p$, the increase of defect energy is difficult to calculate. However, we could estimate an upper bound of the increased defect energy at the contact position as $2\pi\mu \sum_{i=n-9}^{n} r_{0,i}$. The maximum radius of a defect $r_{0,n}$ is the radius of the lens area $r_{lens} = 5$ mm, with which the maximum change of defect energy is $10 \times 2\pi\mu r_{lens} = 1.89 \times 10^{-12}$ N·m, less than ten percent of the elastic energy.

In fact, the radius of the cholesteric sample is usually smaller than the radius of a cylindrical lens. Moreover, circular dislocations only form in a small area that is highly confined due to the special geometry of the crossed cylinders. Outside the confined regime with large heights, defects are network-like without any circular orders. We have shown in previous research [2] that the energy of the shrinking defects is negligible during the retraction and the Derjaguin approximation holds, which is consistent with the study [3] confining smectics in the SFA. Consequently, the experimental results in the past [3-5] and this study were well-fitted with theories without considering defects in similar geometries.

A rigorous description of the dislocation energy may be calculated by simulation in the future to calibrate the data precision.

## III. ANCHORING ENERGY

From Equation B2, the twist angle $\Phi$ is a function of the surface distance $D$,

$$\Phi = \frac{\Phi_0 + \frac{2K_{22}q_0}{W}}{1 + \frac{2K_{22}}{WD}}. \tag{S2}$$

With the parabolic anchoring potential, the free energy can be calculated as,

$$G = \frac{1}{2}K_{22}\frac{(\Phi - q_0 D)^2}{D} + W\left(\frac{\Phi_0 - \Phi}{2}\right)^2 = \frac{1}{2}K_{22}\frac{\left(\frac{\Phi_0 + \frac{2K_{22}q_0}{W}}{1 + \frac{2K_{22}}{WD}} - q_0 D\right)^2}{D} + W\left(\frac{\Phi_0 - \frac{\Phi_0 + \frac{2K_{22}q_0}{W}}{1 + \frac{2K_{22}}{WD}}}{2}\right)^2, \tag{S3}$$

and the corresponding force profile is compared with the profile in the infinite anchoring case, shown in Fig. S4(a). The deviation is small, but indeed, the additional anchoring energy decreases the total forces, and the obtained elastic constant $K_{22} = 4.2$ pN is closer to the molecular property [1].

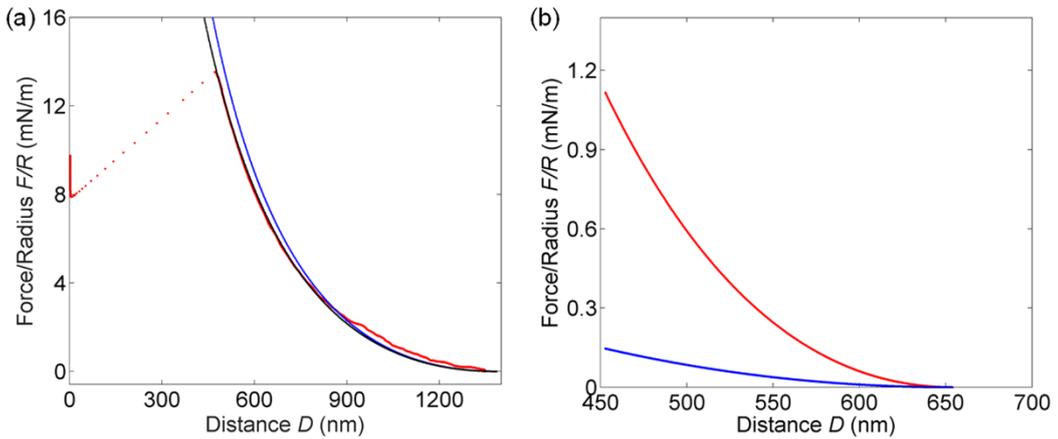

FIG. S4 (a) Force profile (red) in the first regime fitted by elastic forces calculated using Equation A4 (blue line) and Equation S3 (black line) with $K_{22} = 4.2$ pN and $W = 0.15$ mN/m. (b) Elastic forces calculated by Equation A4 (red line) and Equation S3 (blue line) with $K_{22} = 6$ pN and $W = 0.0073$ mN/m.

## IV. SLIDING-SLIP REGIME WITH WEAK ANCHORING STRENGTH

For weak anchoring, the anchoring torque is negligible [6]. Therefore, the elastic torque is mainly balanced by the surface viscosity,

$$K_{22}\left(\frac{\Phi}{D} - q_0\right) = -\gamma_s \frac{\partial \Phi}{\partial t} = -\gamma_s \omega, \tag{S4}$$

$$\frac{\partial D}{\partial t} = v, \tag{S5}$$

$$\frac{\partial \Phi}{\partial t} = \omega, \tag{S6}$$

$$\gamma_s = \eta_s l_s, \tag{S7}$$

where $\eta_s$ is the boundary material viscosity, $l_s$ is the interaction length on the boundary material, $\omega$ is the instant molecular rotation rate on the surface, and $v$ is the instant surface velocity.

From Equations B1 and B2, $\omega = \frac{\partial \Phi}{\partial D} v$, thus Equation S4 becomes,

$$K_{22}\left(\frac{\Phi}{D} - q_0\right) = -\gamma_s \frac{\Phi}{D} v. \tag{S8}$$

In the third regime, the layers are squeezed one by one without large deformation. Therefore, $\Phi \approx \Phi_0 = q_0 D_0$ at a large distance. Equation S8 becomes,

$$K_{22}(D_0 - D) = -\gamma_s v D_0, \tag{S9}$$

$$\gamma_s = -\frac{K_{22}(D_0 - D)}{v D_0}, \tag{S10}$$

$$D = \frac{(K_{22} + \gamma_s v) D_0}{K_{22}}. \tag{S11}$$

The surface viscosity is estimated as $\gamma_s = 1.83 \times 10^{-4}$ Pa s m, using Equation S10 with a typical $K_{22} = 6$ pN, $D_0 = 1000$ nm, $D_0 - D = 61$ nm, and $v = -2$ nm/s. If the interaction length on boundary materials is around 10 nm, then the boundary material viscosity $\eta_s = 1.83 \times 10^4$ Pa s, which is very large.

The surface viscous torque in Equation S8 is integrated over $\Phi$ to calculate the surface viscous free energy $G_{sv}$ and is substituted by Equation S10,

$$G_{sv} = -\gamma_s \frac{\Phi^2}{2D} v \approx \frac{1}{2} K_{22} q_0^2 \frac{(D_0 - D) D_0}{D}. \tag{S12}$$

The force induced by surface viscosity is estimated as $F = 0.8$ mN/m at $D_0 = 1000$ nm, which is very close to the background force in the third regime in Fig. 1(b). As a result, surface viscosity stretches cholesteric layers and the defects (Movie S1), in addition to the regular elastic forces.

For the first two regimes, the anchoring is strong enough to expel the defects to the center [7]. Therefore, there is no interaction between surfaces and defects. However, in the third regime, weak anchoring attracts defects to the wall, generating binding energy [8-12] in addition to the surface torque.

In some studies [8-12] with smectics confined in the SFA, defects also escaped to the surfaces under homeotropic anchoring, which is typically weaker than planar anchoring. Given that the critical anchoring $2\sqrt{\frac{3}{8} K_{33} B}$ for surfaces to expel defects to the bisector plane [7] is larger for smectics than for cholesterics, defects in smectics were pinned on the surfaces. Those studies [10-12] explained the avalanche events by the stretch of screw dislocations pinned on surfaces. However, for cholesterics with larger layers compared to smectics, defects are present in smaller amounts. The pinned defects cannot be the only viscous force contributing to the large background force. The sliding of whole cholesteric planes must also be taken into consideration. This may shed light on the commonly observed background forces in past studies [13,14] with liquid crystals.

## V. SIMULATION WITH A 4X4 MATRIX

Simulation with a 4x4 matrix has been used in the past [2,15] to reproduce the optics formed in the SFA. Based on the twist angle profile in Fig. 4(c), the simulations are compared by subtracting or adding one or two layers, as shown in Fig. S5. Overall, the simulation changes significantly after one layer is subtracted, but makes negligible differences by adding one or two layers during the retraction. With the addition of layers, the simulation is not sensitive to the change in twist angle because the layers are largely compressed, which is similar to the isotropic limit [15]. This indicates that with multiple twisted layers, the 4x4 matrices may not accurately reflect the correct twist angles.

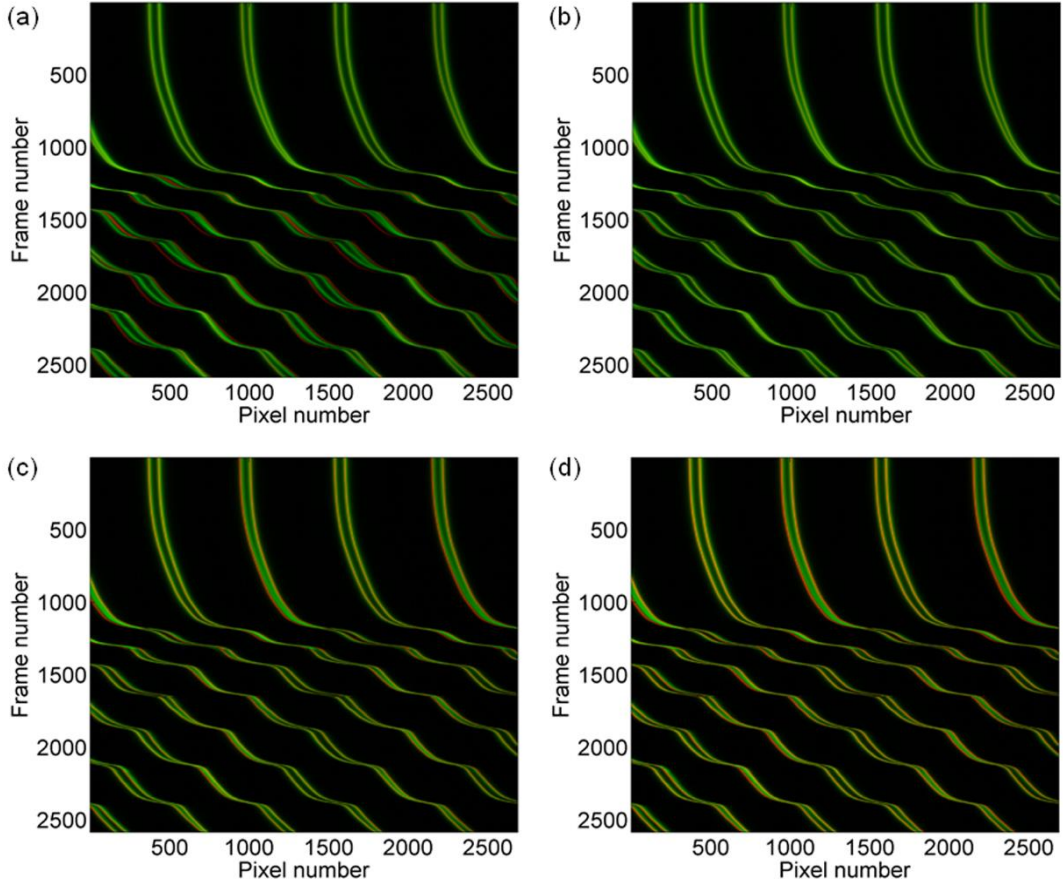

FIG. S5 Overlay of the simulation and experiment spectrograms during retraction in the third regime using ImageJ. The intersection angle between two mica surfaces is 57.4 °. The twist angle in the simulation is the corresponding twist angle profile in Fig. 4(c), (a) by subtracting $\pi$, (b) with no change, (c) by adding $\pi$, and (d) by adding $2\pi$. The frame number is proportional to the time, the thick green line is the experimental data, the thin red line is the simulation data, and the overlaid region shows yellow.

Movie S1 The asymmetrical stretch of a defect on the surface during retraction in the third regime. Fringes of equal chromatic order, formed by multiple-beam interferometry, were observed on the spectrometer.